\documentclass[showpacs,aps,twocolumn,prb,groupedaddress]{revtex4-1}

\usepackage{graphicx}
\usepackage{color}
\usepackage{amsmath}
\usepackage{verbatim}
\usepackage{amssymb,ulem}
\renewcommand{\emph}[1]{\textit{#1}}

\newcommand{\vek}[1]{\underline{#1}}
\newcommand{\vk}{\underline{k}}
\newcommand {\om}  {\ensuremath{\omega}}

\newcommand{\eps}{\epsilon}
\newcommand {\cd}  {\ensuremath{\hat c^{\dagger}}}
\renewcommand {\c}  {\ensuremath{\hat c^{\phantom{\dagger}}}}
\newcommand {\ull}[1] {\ensuremath{\underline{\underline{#1}}}} 

\newcommand{\DMFTTS}{NLD} 

\begin{document}

\title{Extension of dynamical mean-field theory by  inclusion of nonlocal 
  two-site correlations with variable distance}

\author{Torben Jabben and Norbert Grewe}
\affiliation{Institut f\"ur Festk\"orperphysik, Technische Universit\"at Darmstadt, 
  Hochschulstr.\ 8,  64289 Darmstadt, Germany}
\author{Sebastian Schmitt}
\email[Email: ]{sebastian.schmitt@tu-dortmund.de}
\affiliation{Lehrstuhl f\"ur Theoretische Physik II, Technische Universit\"at Dortmund,
  Otto-Hahn-Str.\ 4, 44221 Dortmund,  Germany}

\date{\today}

\begin{abstract}
  We present a novel approximation scheme for the 
  treatment of strongly correlated electrons in arbitrary 
  crystal lattices. The approach extends the well-known
  dynamical mean field theory  to include  nonlocal 
  two-site correlations of arbitrary spatial extent.
  We extract the nonlocal correlation functions from  two-impurity
  Anderson models where the impurity-impurity distance
  defines the spatial extent of the correlations included.
  Translational invariance is
  fully respected by our approach since correlation functions of any
  two-impurity cluster
  are periodically embedded to $\vek k$-space via a Fourier transform.
  As a first application, we study the two-dimensional Hubbard model
  on a simple-cubic lattice. We demonstrate how pseudogap formation
  in the many-body resonance at the Fermi level results from the inclusion of 
  nonlocal correlations. 
  A comparison of the spectral function with  the dynamical-cluster 
  approximation shows qualitative agreement of high- as well as low-energy 
  features.
\end{abstract}



\maketitle

\section{Introduction}
Compounds with strongly correlated electrons\cite{colemanUnfinishedFrontier03,*quintanillaStrongCorr09,ramakrishnanSCES08}
are in the focus of modern solid-state research.
They usually contain transition metal ions where part of the
valence electrons move in narrow bands or are nearly localized. 
Prominent  examples 
include  high-temperature cuprate superconductors,\cite{plakidaHighTCBook10}
frustrated  magnets\cite{lacroixFrust2010} and heavy-fermion 
systems.\cite{greweSteglichHF91,*colemanHeavyFermionMag07} 

Despite a large effort, various fundamental questions are still 
open.\cite{normanPseudogap05,vojtaMobileItinernantCuprate09,greweSteglichHF91,*colemanHeavyFermionMag07,ramakrishnanSCES08}
The model system of a single transition metal ion immersed into a 
metallic matrix has essentially been solved by 
analytical\cite{wiegmannExactSIAMI83,*tsvelickExactSIAMII83,*andreiSolutionKondo83,hewson:KondoProblem93} and 
powerful numerical methods.\cite{wilsonNRG75,*bullaNRGReview08,*hirschQMC86,*wernerContinousTimeQMC06,*whiteDMRG92,greweCA108}
However, the more general case of regular crystal
lattices with strong electron correlations in or near a metallic
regime is much more difficult to handle. The large number
of relevant electronic states and the interplay
of hybridization and interaction effects makes it even
difficult to identify relevant degrees of freedom or 
arrive at low-energy effective models.

An understanding of electronic correlations lies at the heart of 
most open problems in this field. Although the metallic character of many materials 
lends hope to the hypothesis that relevant interactions are screened and
thus of short range, they can in principle drive correlations over all
ranges of spatial distances.
 
Modern approaches\cite{grewe:spectrumPAM84,*Grewe:lnca83,*grewe:LNCA87,*kuramoto:XNCA85} 
focused on the dominant role of the local Coulomb repulsion. 
In particular, dynamical mean-field 
theory\cite{pruschke:dmftNCA_HM95,*georges:dmft96,*georgesDMFTreview04,*vollhardtDMFT11} (DMFT) has proven 
to be very successful. It is capable 
of describing  lattice versions of the Kondo effect in a regime of
sizable  local magnetic moments and provides valuable insights to the 
correlation driven Mott-Hubbard metal-to-insulator transition.\cite{pruschkeMITTwoOrbHM06,*sordiMITPAM07,*held:MottTransPAM_neighHyb00,*pruschkeLowEnergiePAM00,*bulla:MITHM99}
These approximations have met with considerable success concerning,
for example, one-particle properties\cite{PhysRevB.58.3690,*kotliarDMFT04,*maiti:165128,*greteKink11,*schmittNFLvHove10,grewe:quasipart05} 
or susceptibilities.\cite{zlatic2PVertexDMFT90,*jarrellQMCHM92,*jarrell:symmetricPAM95,*kunesTwoDMFT11,*schmittPhD08,*schmittMagnet11} 

However, many phenomena require the inclusion of  nonlocal correlations. 
Prominent examples are systems near a quantum 
critical point\cite{stewartNFLReview01,*stewartNFLReview01app06,*siQCHF10,*loehneysenFLQCP07,*sachdevQCP10,*brounQCP08}
or cuprate high-temperature 
superconductors.\cite{plakidaHighTCBook10}
In recent years extensions of DMFT have been put forward to
remedy these shortcomings by the inclusion of some spatial
correlations between electrons.\cite{maier:QuantumCluster05,*smithSpatialCorrDMFT00,*toschiDGA07,*kataninDGA2dHM09,*rubtsovDualFermion08,*rubtsovDual09,*tremblayTPSC11} 
In the cluster approaches, the 
problem is mapped to an effective cluster of few lattice sites and,
in analogy to DMFT, this cluster is treated like a complex impurity in a 
dynamic external field. 

These approaches capture short-range correlations quite accurately
and have contributed considerably to the understanding of pseudogaps and 
shadow bands,\cite{huscroft:pseudogapHM01,*moukouriMIT01,*jarrell:QMCMEMNonLocalDMFT01,vilk:ShadowCuprate97,*Scalapino2006,*kyungPseudogap06,*tremblayPseudogap06,normanPseudogap05,macridinPseudogapHM06}
as well as to the possible occurrence of quantum critical points.\cite{vidhyadhiraja2DHMNFL09,*khatamiQCPHM10,*yangQCSC11,chenVanHove11}

However, some  problems remain in these theories.
In cluster DMFT\cite{kotliarCDMFT01,*biroli:CDMFTCausality04} (CDMFT), for example, 
the translational invariance of the  crystal lattice is not 
fully respected. The dynamical 
cluster approximation\cite{hettler:NonLocalDCA98,*hettlerDCA00} (DCA)
remedies this shortcoming, but the question remains  
which choice for the size and geometry of the cluster is advantageous.\cite{isidoriCluster09}
But the most fundamental limitation is the restriction to 
rather short-ranged correlations.

In this paper we present a novel kind of self-consistent 
approach to correlated lattice systems which is in principle
capable of including nonlocal correlations of arbitrary distance.
It extends the well-known DMFT by the inclusion of two-site correlations
of all length-scales and is thus termed \textit{nonlocal DMFT} (\DMFTTS{}).
These correlations are incorporated by
a mapping of the lattice model onto a multitude of 
two-impurity Anderson models (TIAM),
where the impurity-impurity distance is  varied.
The translational invariance and crystal symmetries are fully respected by
the  scheme proposed here. 

The details of the mapping between lattice and two-impurity models
is presented in Sec.~\ref{sec:theory}.
As a first application we study in Sec.~\ref{sec:application} 
the two-dimensional Hubbard model  on a square lattice, a model which 
is usually considered in connection with cuprate superconductors.
We utilize the recently developed two-impurity enhanced noncrossing 
approximation\cite{jabbenENCATI12} as the two-impurity solver
and analyze the results obtained with
respect to the quality of our novel lattice approach and to
their physical implications.
A short conclusion and outlook is given in Sec.~\ref{sec:conclusion}.

\section{Self-consistent lattice theory}
\label{sec:theory}

In order to introduce the ideas underlying our self-consistent
scheme we restrict the 
following discussion to the example of a single-band Hubbard model
\begin{align}
  \label{eq:2}
  \hat H=& \sum_{\vk,\sigma} \left(\epsilon +t_{\vk}\right)\cd_{\vk,\sigma} \c_{\vk,\sigma}
  + U\sum_{j}
  \hat n_{j,\uparrow}\hat n_{j,\downarrow}.
\end{align}
We use a mixed representation, where the kinetic energy is expressed
in terms of band states with momentum $\vk$ and spin $\sigma=\{+,-\}=\{\uparrow,\downarrow\}$ 
which 
are created (annihilated) by the  operators $\cd_{\vk,\sigma}$
($\c_{\vk,\sigma}$). $\epsilon$ denotes the local single-particle
energy and the dispersion $t_{\vk}$ is 
the Fourier transformation of the hopping matrix elements $t_{ij}$ between lattice 
sites $i$ and $j$ with lattice vectors $\vek{R}_i$ and $\vek{R}_j$, respectively.  
The lattice structure, i.e.\ the type of lattice and topology of the 
hopping matrix elements, is completely encoded into the dispersion 
relation $t_{\vk}$.
The interaction energy characterized by the Coulomb matrix 
element $U$ is conveniently written
in terms of local occupation number operators $\hat n_{j,\sigma} =\cd_{j,\sigma}\c_{j,\sigma}$
which measure an electron on site $j$ with spin $\sigma$.
The lattice constant $a=1$ will serve as the fundamental length scale
and the spin index $\sigma$ will be dropped whenever possible
without ambiguities. 

The extension of the  novel approach presented here to  more general systems
with multiple bands and/or more complicated interactions can be readily
obtained.

\begin{figure}[t]
  \centering 
  \includegraphics[width=0.9\linewidth]{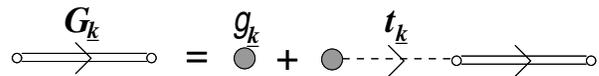}
  \caption{Diagrammatic representation of Eq.~\eqref{eq:4}.
  }
  \label{fig:1}
\end{figure}

The fundamental quantity of interest is the one-particle Green function
which obeys the Dyson equation
\begin{align}
  \label{eq:4}
  G_{\vek{k}}(z)&=g_{\vek{k}}(z)+g_{\vek{k}}(z)t_{\vek{k}}(z) G_{\vek{k}}(z)
\end{align}
which is represented graphically in Fig.\ \ref{fig:1}.
The formal solution is given by 
\begin{align}
  \label{eq:eq1}
  G_{\vek{k}}(z)&=\left[g_{\vek{k}}(z)^{-1}-t_{\vek{k}}\right]^{-1}
  .
\end{align}
The correlated part 
\begin{align}
  \label{eq:eq1b}
  g_{\vek{k}}(z)&=[z-\epsilon-\Sigma_{\vek{k}}(z)]^{-1}
\end{align}
incorporates  the irreducible self-energy $\Sigma_{\vk}(z)$
and accounts for the correction to the noninteracting 
system due to the interaction term proportional to $U$ 
in the Hamiltonian Eq.~\eqref{eq:2}. It describes the 
interaction-induced, i.e.\ correlated, part of the  propagation process, in which single-particle transfers 
via elementary hoppings are excluded.

It is useful to re-write the Dyson equation \eqref{eq:4} in terms of the
lattice $T$-matrix  $T_{\vek{k}}$,
\begin{align}
  \label{eq:4b}
  G_{\vek{k}}(z)
  &=g_{\vek{k}}(z)+g_{\vek{k}}(z)T_{\vek{k}}(z) g_{\vek{k}}(z)
  ,\end{align}
 which is commonly defined via
 \begin{align}
  \label{eq:Tka}
  T_{\vek{k}}(z) &=t_{\vek{k}}+t_{\vek{k}}G_{\vek{k}}(z) t_{\vek{k}}
  \\ \label{eq:Tkb}
  &=\frac{t_{\vek{k}}}{1-g_{\vek{k}}(z) t_{\vek{k}}}
  .
\end{align}
In the last equality the explicit form of Eq.~\eqref{eq:eq1} is used.

Processes contributing to the irreducible self-energy $\Sigma_{\vek{k}}(z)$
can be grouped together according to the nature and degree of correlations  
they contain. One contribution represents the exact solution of an
isolated interacting local site, which is represented  by the 
``atomic'' self-energy\cite{hubbard:hI63} 
$\tilde{\Sigma}^{(0)}(z)=U\frac{(z-\epsilon)(1-\langle \hat n_\sigma\rangle)}{z-\epsilon-U\langle \hat n_\sigma\rangle}$.
All other terms incorporate genuine lattice processes. Some of these are 
captured by the well-known self-energy of the DMFT-approximation,
which furnishes a $\vek k$-independent but dynamic contribution $\tilde{\Sigma}^{(1)}(z)$.
It incorporates loops through the lattice attached at 
one site at which all local (dynamic) correlations are fully respected.   
Beyond DMFT, correlations between two or more of these loops are generated by 
interaction events
at different sites  and thus constitute nonlocal cumulant corrections.

Organizing these cumulant corrections according to the number of different 
lattice sites they correlate,
the correlation function 
of Eq.~\eqref{eq:eq1b} can be expanded as
\begin{align}
  \label{eq:3}
  g_{\vek{k}}^{-1}(z)=&z-\epsilon-\tilde{\Sigma}^{(0)}(z)-\tilde{\Sigma}^{(1)}(z)
  \\\notag
  & -\tilde{\Sigma}^{(2)}_{\vk}(z)-\tilde{\Sigma}^{(3)}_{\vk}(z)-\ldots 
\end{align}
It is the aim of the present investigation to identify and calculate the
self-energy contribution  $\tilde{\Sigma}^{(2)}_{\vk}(z)$, which contains 
all  nonlocal correlations  between any two sites of the lattice.
While two-site correlations are thus explicitly included in this scheme,
three-site and higher-order nonlocal cumulant corrections $\tilde{\Sigma}^{(n)}_{\vk}(z)$
with $n\geq 3$ 
are neglected. 
The function incorporating these correlations, $g_{\vek k}(z)$,
will be extracted from solutions of various two-impurity models where 
the impurity-impurity distance is varied. Then, the Dyson 
equation \eqref{eq:4}, respectively its solution Eq.~\eqref{eq:eq1},
is used to obtain the lattice Green function.   

\subsection{Mapping between the lattice and a set of two-impurity models}

In order to extract the function $g_{\vek k}(z)$ which includes two-site 
correlations, we consider a two-impurity Anderson model\cite{PhysRev.133.A1594} 
(TIAM). The two magnetic impurities are immersed with finite distance
into a host with a noninteracting conduction band.  
The Hamiltonian of one such general  TIAM with a fixed distance vector $\vek a$ 
is given by
\begin{align}
  \label{eq:2ImpHam}
  \hat H^{(\mathrm{2imp},\vek a)}=&
  \sum_{j=\{1,2\},\sigma} \epsilon_{j}\hat c^\dagger_{j,\sigma}  \hat c_{j,\sigma}^{\phantom{\dagger}}
  + \sum_{j,l,\sigma,\sigma'}U^{(\vek a)}_{jl}
  \hat n_{j,\sigma}
  \hat n_{l,\sigma'}
  \\\notag &
  +\sum_{j,\sigma}t^{(\vek a)}_{jl} \hat c^\dagger_{j,\sigma}  \hat c_{l,\sigma}^{\phantom{\dagger}}  
  + \sum_{j,l}\hat W^{(\vek a)}_{jl}
  \\ \notag  &
  +\sum_{\vk,\sigma} \epsilon_{\vk}\hat a^\dagger_{\vk,\sigma}  \hat a_{\vk,\sigma}^{\phantom{\dagger}}
  \\\notag
  &+\frac{1}{\sqrt{N}}\sum_{\vk,j,\sigma} \big( V_{\vek k}e^{-i\vek k\, \vek R_j}
  \hat a^\dagger_{\vk,\sigma}  \hat c_{j,\sigma}^{\phantom{\dagger}}+h.c.\big)
  .
\end{align}
The $\hat c$-operators  ($\hat a$-operators) describe local interacting electrons  
on the impurities at positions $\vek R_j\in \{\vek R_1,\vek R_2\}$ 
(the noninteracting conduction 
band electrons with momentum $\vek k$)
with spin $\sigma$. The spatial distance between the two impurities is 
fixed and given by $\vek{a}=\vek R_1-\vek R_2$. $\epsilon_j$ is the local single-particle
energy on each impurity site and $U^{(\vek a)}_{lj}$ are the
Coulomb matrix elements of
the local ($j=l$) and nonlocal ($j\neq l$) density-density interactions.
The term proportional to $t^{(\vek a)}_{jl}$ is a possible direct single-particle
hopping inside the two-impurity cluster and $\hat W^{(\vek a)}_{lj}$ collects 
additional interactions as, for example,  exchange, pair hopping or 
correlated hopping. The last term represents the  hybridization between 
interacting electrons of the two-site cluster and noninteracting band 
electrons. 

The scheme proposed  is capable of properly treating all such terms
in the Hamiltonian so that nonlocal Coulomb interaction or long-ranged single-particle
hopping matrix elements can in principle be considered. For the present 
case of the single-band Hubbard model nonlocal Coulomb matrix elements 
are not included and the single-particle
hopping coincides with that of the underlying lattice as specified in the Hamiltonian
of Eq.~\eqref{eq:2}. 

Since we are only interested in properties of the interacting $c$-electrons
neither the hybridization matrix elements nor the dispersion 
$\epsilon_{\vk}$ of the  band electrons are directly of interest. 
Only the effective medium for the two-impurity cluster is relevant
which is expressed in terms of these  
parameters as
\begin{align}
\label{eq:tt}
  \tilde{T}^{(2\mathrm{imp}, \vek a)}_{\vek a}(z)&=\frac 1 N \sum_{\vek k} e^{-i\vek k\, \vek a} \frac { |V_{\vek k}|^2}{z-\epsilon_{\vek k}}
  .
\end{align}
It represents propagation processes via the  
noninteracting conduction band in form of  irreducible loops
starting and ending at the two-site cluster.
In our truly  self-consistent theory for a lattice model a corresponding 
effective medium $\tilde{T}^{(2\mathrm{imp},\vek a)}_{\vek a}$ has to be determined 
by an appropriate mapping between the lattice and the two-impurity model.  
Therefore, it becomes dressed and takes into account repeated 
interactions on local shells of other lattice sites.

A Dyson equation for the local two-impurity Green functions
is set up as follows
\begin{align}
  \label{eq:11}
  \ull{G}^{(2\mathrm{imp},\vek{a})}&=\ull{g}^{(2\mathrm{imp},\vek{a})}
  \\\notag 
  & +\ull{g}^{(2\mathrm{imp},\vek{a})} \left(\ull{t}^{(\vek a)} +\ull{\tilde T}^{(2\mathrm{imp},\vek{a})}\right) \ull{G}^{(2\mathrm{imp},\vek{a})}
  ,
\end{align}
where we used a  matrix notation for the spatial components
\begin{align}
  \label{eq:matrixNot}
  \ull{A}^{(2\mathrm{imp},\vek{a})}=
  \left(
    \begin{matrix}
      A^{(2\mathrm{imp},\vek{a})}_{\vek{0}} & A^{(2\mathrm{imp},\vek{a})}_{\vek{a}}\\[2mm]
      A^{(2\mathrm{imp},\vek{a})}_{-\vek{a}} & A^{(2\mathrm{imp},\vek{a})}_{\vek{0}}
    \end{matrix}\right)
  .
\end{align}
The correlated part $\ull{g}^{(2\mathrm{imp},\vek{a})}$ represents  the cluster cumulant
Green function and takes the interaction matrix elements into account. 
Therefore, the  direct $\sim \ull{t}^{(\vek a)}$
and indirect hopping events via the effective medium $\sim \ull{\tilde T}^{(2\mathrm{imp},\vek{a})}$
are explicitly incorporated in Eq~\eqref{eq:11}. 

The Dyson equation \eqref{eq:11} can be formally solved to yield
\begin{equation}
  \label{eq:14}
  \ull{G}^{(2\mathrm{imp},\vek{a})}(z)=\left[{\ull{g}}^{(2\mathrm{imp},\vek{a})}(z)^{-1}-\ull{t}^{(\vek{a})}-\tilde{\ull{T}}^{(2\mathrm{imp},\vek{a})}(z)\right]^{-1}
  ,
\end{equation}
which apart from the matrix structure and the occurrence of the hopping matrix
$\ull{t}^{(\vek{a})}$ has the same form as in the single-impurity case.\cite{hewson:KondoProblem93}

It is instructive to formulate the above equations in terms
of the full $T$-matrix of the TIAM, thereby establishing an equivalence to
the lattice equations \eqref{eq:4b} to \eqref{eq:Tkb}. 
The local Green function can be expressed as 
\begin{equation}
  \label{eq:Tmat2Imp}
  \ull{G}^{(2\mathrm{imp},\vek{a})}=
  \ull{g}^{(2\mathrm{imp},\vek{a})}
  +\ull{g}^{(2\mathrm{imp},\vek{a})} \ull{T}^{(2\mathrm{imp},\vek{a})} \ull{g}^{(2\mathrm{imp},\vek{a})}  
  .
\end{equation}
In contrast to the irreducible medium $\tilde{\ull{T}}^{(2\mathrm{imp},\vek{a})}$, 
the $T$-matrix  incorporates repeated visits of the two-impurity cluster. 
The $T$-matrix is thus built up from irreducible loops, accounted for by the inclusion 
of $\ull{t}^{(\vek a)}+\ull{\tilde{T}}^{(2\mathrm{imp},\vek{a})}$
for steps inside and outside the cluster,
and repeated dwellings inside the cluster, where each visit contributes a 
factor  $\ull{g}^{(2\mathrm{imp},\vek{a})}$.
It can be  expressed as 
\begin{align}
  \label{eq:12}
  \ull{T}^{(2\mathrm{imp},\vek{a})}&=\left[\left(\ull{t}^{(\vek a)}+\ull{\tilde{T}}^{(2\mathrm{imp},\vek{a})}\right)^{-1}-\ull{g}^{(2\mathrm{imp},\vek{a})}\right]^{-1}
  .
\end{align}

A mapping between the two-impurity and the lattice model is 
accomplished by connecting the irreducible correlation functions $g_{\vek{a}}(z)$ 
of both approaches. 
For each nonzero distance vector $\vek a$  with solution $g^{(2\mathrm{imp},\vek{a})}_{\vek{a}}(z)$ of
the two-impurity model,
we identify  
\begin{align}
  \label{eq:gaMap}
  g_{\vek{a}}(z)&\overset ! =\frac 1{\nu_{a}}  g^{(2\mathrm{imp},\vek{a})}_{\vek{a}}(z) 
  \qquad (\vek a\neq \vek 0)
  \intertext{where the matrix elements [cf.\ Eq.~\eqref{eq:matrixNot}]}
  g_{\vek{a}}(z)&=\frac 1N\sum_{\vek k}e^{-i\vek k\,\vek a}g_{\vek{k}}(z)
\end{align}
is the  Fourier transform of the lattice correlation function of Eq.~\eqref{eq:eq1b} and
$\nu_{a}$ the number sites which have a distance $a=|\vek{a}|=\sqrt{\sum_i a_i^2}$
to a given site. 

The prefactor of $\frac 1{\nu_{a}}$ in Eq.~\eqref{eq:gaMap} arises due to the 
restricted phase-space for scattering in the two-impurity model compared to the
lattice situation. In the two-impurity cluster the interaction-induced 
scattering always transfers the electron from one site to the other at distance $\vek a$.
In contrast, in the lattice the scattering process transfers an electron only in a fraction
$\frac 1{\nu_{a}}$ of cases to  one specific site with distance $a=|\vek a|$.

There is some ambiguity for the correlation function with zero distance $\vek a=\vek 0$.
While in the lattice there is only one such function,
$g_{\vek 0}(z)$, there exist many such functions $g_{\vek 0}^{(2\mathrm{imp},\vek{a})}$
from all the effective two-impurity models, one for each distance $\vek a$.
Then the question arises, which of all these functions should be chosen. 
To this end, we take a constructive approach, where we start from a reference correlation
function  $g^{\mathrm{DMFT}}(z)$ obtained
with the  DMFT and add to it all additional correlations of representative  
two-impurity clusters at different distances
\begin{align}
  \label{eq:g0Avg}
  g_{\vek 0}(z)=&g^{\mathrm{DMFT}}(z) \\
  & \notag 
  +\sum_{|\vek a|} \left[g_{\vek 0}^{(2\mathrm{imp},\vek{a})}(z)-g^{\mathrm{DMFT}}(z)\right]
  .
\end{align}

The momentum dependent correlation function of the lattice theory 
is now given by the  inverse Fourier transform  
\begin{align}
  \label{eq:gek}
  g_{\vek{k}}(z)&=\sum_{\vek a}e^{i\vek k\,\vek a}g_{\vek{a}}(z)
\end{align}
which directly leads to the full lattice Green function via Eq.~\eqref{eq:eq1}.

As explained after Eq~\eqref{eq:tt}, the effective medium 
$\ull{\tilde T}^{(2\mathrm{imp},\vek a)}$  in the framework of our self-consistent
theory has to be obtained from lattice quantities.  
This is achieved by identifying the momentum dependent $T$-matrices of both
models,
\begin{align}
T_{\vek k}(z)\overset !=   T^{(2\mathrm{imp})}_{\vek k}(z)=\sum_{\vek a}e^{i\vek k\,\vek a} \,T^{(2\mathrm{imp},\vek a)}_{\vek a}(z) .
\end{align}
The momentum dependent $T$-matrix $T_{\vek k}$ incorporates the 
correct translationally invariant linear combination of two-site propagations
with a fixed distance.
Inverting the Fourier transform and using Eq.~\eqref{eq:Tkb} we
obtain the $T$-matrix  for a  two-impurity model with fixed
distance $\vek a$
\begin{align}
\label{eq:2impT}
  T^{(2\mathrm{imp},\vek a)}_{\vek a}=\frac 1N\sum_{\vek k}e^{-i\vek k\,\vek a}\frac{t_{\vek{k}}}{1-g_{\vek{k}}(z) t_{\vek{k}}}
  .
\end{align}
From this  the irreducible medium for the two impurity model can be obtained   by inverting Eq.~\eqref{eq:12}
and inserting the known expression for $\ull{g}^{(2\mathrm{imp},\vek{a})}$.

Lattice and crystal symmetries are completely 
respected in the  present approach. Lattice-translational invariance is
incorporated by construction. Even though the two-impurity 
models are solved  for distances  in  real-space, the correlation
functions $g^{(2\mathrm{imp},\vek a)}_{\vek{a}}(z)$, which explicitly exclude 
single-particle hopping, can be periodized for this purpose with the help of
a Fourier transform
to yield the lattice correlation function  $g_{\vek{k}}(z)$.
Crystal symmetries are also fully respected since all the two-impurity
models for which the distance vectors $\vek a_i$ are generated by 
point-group transformations,  have identical 
effective media and are therefore identical.
Point group symmetries can even be used to decrease the computational 
cost as only one representative two-impurity model 
of such a group of symmetry-related points needs to be solved. 
The solutions of all the others follow by symmetry operations.

We conclude this section by summarizing the steps of
the calculation scheme:
\begin{enumerate}
\item Chose a set of $N_a$ representative distance vectors
  $\{\vek{a}_i\}$ with $i=1,\ldots,N_a$, one for each 
  set of symmetry-related distance vectors.
\item Start with an initial guess for the effective media  $\tilde{\ull{T}}^{(2\mathrm{imp},\vek{a}_i)}$
  for each distance vector.
\item \label{item}
  Solve $N_a$ different effective two-impurity models, one for each representative  distance $\vek{a}_i$
  with the  medium $\tilde{\ull{T}}^{(2\mathrm{imp},\vek{a}_i)}$.
  The result are $N_a$ (matrix) Green functions $\ull{G}^{(2\mathrm{imp},\vek{a}_i)}$
  from which the correlation functions $\ull{g}^{(2\mathrm{imp},\vek{a}_i)}$
  are derived via the inversion of Eq.~\eqref{eq:14},
  \begin{align}
    \ull{g}^{(2\mathrm{imp},\vek{a})}=\left[{\ull{G}}^{(2\mathrm{imp},\vek{a})}{}^{-1}+\ull{t}^{(\vek{a})}+\tilde{\ull{T}}^{(2\mathrm{imp},\vek{a})}\right]^{-1}.
  \end{align}
  
\item Map the correlation functions of the two-impurity models to their lattice counterparts 
  for fixed distances with Eqs.~\eqref{eq:gaMap} and \eqref{eq:g0Avg}. 
  Fourier-transform these $g_{\vek{a}_i}(z)$ via Eq.~\eqref{eq:gek}
  to obtain $g_{\vek k}(z)$.

\item Use $g_{\vek k}(z)$ and Eq.~\eqref{eq:2impT} to get the two-impurity 
  $T$-matrix  $T^{(2\mathrm{imp},\vek a_i)}_{\vek a_i}$ for each distance 
  vector $\vek a_i$.
  A new guess for the effective medium
  $\tilde{\ull{T}}^{(2\mathrm{imp},\vek{a}_i)}$ is obtained with the 
  inversion of   Eq.~\eqref{eq:12},
  \begin{align}
    \label{eq:21}
    \ull{t}^{(\vek a)}+\ull{\tilde{T}}^{(2\mathrm{imp},\vek{a})}&=\left[{\ull{T}^{(2\mathrm{imp},\vek{a})}}^{-1}+\ull{g}^{(2\mathrm{imp},\vek{a})}\right]^{-1}
    .
  \end{align} 
\item  Go back to step \ref{item} and iterate until convergence is
  reached.
\end{enumerate}

One-particle spectra of the lattice are then obtained from 
Eq.~\eqref{eq:eq1} in momentum space and the local function
via a Fourier transform. 

An earlier approach of Schiller and Ingersent 
also utilizes a two-impurity model to extend the DMFT 
to include nonlocal correlations.\cite{schiller:TwoImpurity95}
However there are various differences to our approach. 
They consider an expansion in the inverse spatial dimension $\frac 1d$ 
and consequently employ the two-impurity model only
for one specific distance, i.e.\  nearest-neighbor sites only.
Additionally, they identify the irreducible self-energies [cf.\ Eq.~\eqref{eq:eq1b}]
of the lattice and the two-impurity model.  
In contrast, we establish the mapping between the lattice and the effective impurity models
via the correlation function $g_{\vek a}$.
This represents a crucial difference, since $g_{\vek{a}}$ includes \textit{all} 
propagation processes correlated between the two sites via the interaction. Thus, it also incorporates 
repeated interaction-induced scattering between two sites which are excluded from 
the irreducible self-energy but need to be accounted for in the lattice Dyson 
equation \eqref{eq:4}.
An important point concerns the incorporation of the explicit 
hopping between nearest neighbor sites $t^{(\vek a)}$. 
In the present approach the single-particle transfers and free propagations 
through the medium are explicitly separated from correlation effects in both 
models, cf.\ Eqs.~\eqref{eq:4} and \eqref{eq:11}.
This implies for the two-impurity model the hopping to be incorporated into the effective 
medium [see  Eqs.~\eqref{eq:11} and \eqref{eq:12}] and  ensures translational invariance.
Additionally, any imbalance in the 
treatment of inter- and intra-cluster hopping in the two-impurity model,
as it occurs, for instance, in the CDMFT,   is removed.




\section{self-consistent scheme applied to the 2$d$-Hubbard model}
\label{sec:application}

In this section we present results for the two-dimensional
($d=2$) Hubbard model \eqref{eq:2} obtained with the approximation 
scheme described in the previous section. 
We focus on the metallic regime at not too low temperatures.
Long-ranged magnetic order and superconducting states are thus
excluded, whereas magnetic correlations of finite extent and the
precursor regime of a metal-insulator transition are  included 
and accessible.

For the solution of the two-impurity problems we employ a solver 
based on direct perturbation theory 
in the hybridization\cite{greweCA108,keiter:PerturbPRL70,*grewe:IVperturb81,*keiterMorandiResolventPerturb84} 
which is an extension of the two-orbital
solver\cite{greweENCACF09} and which is described 
elsewhere.\cite{Jabben2010,jabbenENCATI12} In contrast to the 
usual noncrossing approximation\cite{Grewe:siam83,*Kuramoto:ncaI83,*bickers:nca87a}
(NCA) this two-impurity enhanced
noncrossing approximation  includes vertex 
corrections which allow for the accurate description of finite 
Coulomb repulsions.

An important requirement for the two-impurity solver is, it needs to 
be able to treat dynamic non-diagonal effective media 
$\tilde{T}^{(2\mathrm{imp},\vek{a})}_{\vek a}(z)$.

\begin{figure}[ht]
  \includegraphics[width=0.8\linewidth]{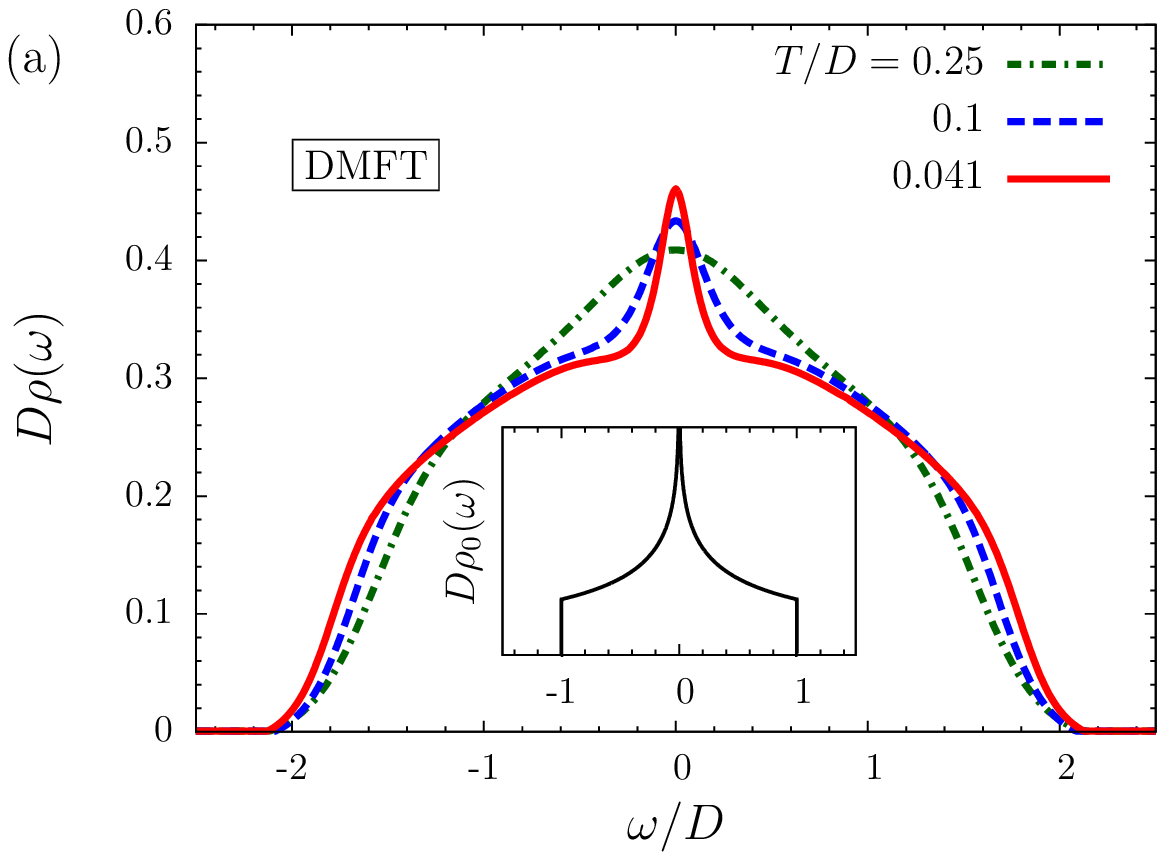}
  \includegraphics[width=0.8\linewidth]{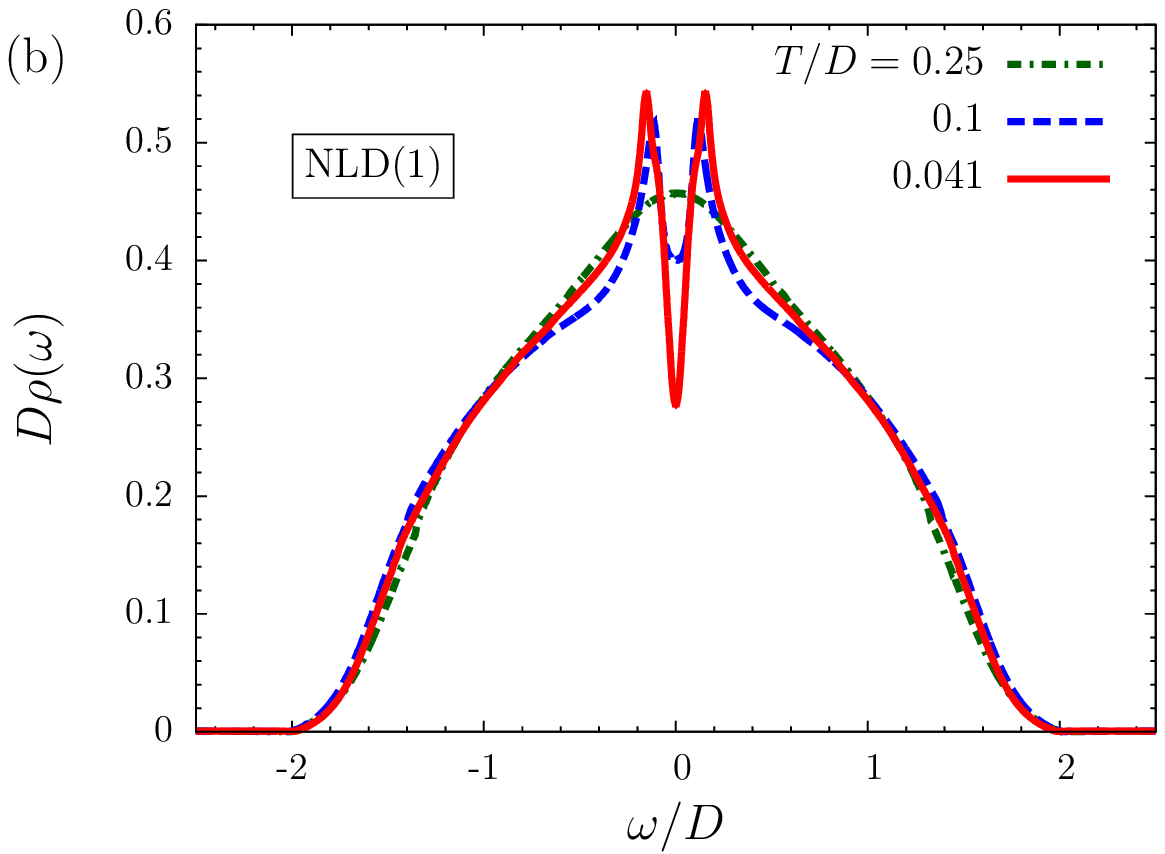}
  \includegraphics[width=0.8\linewidth]{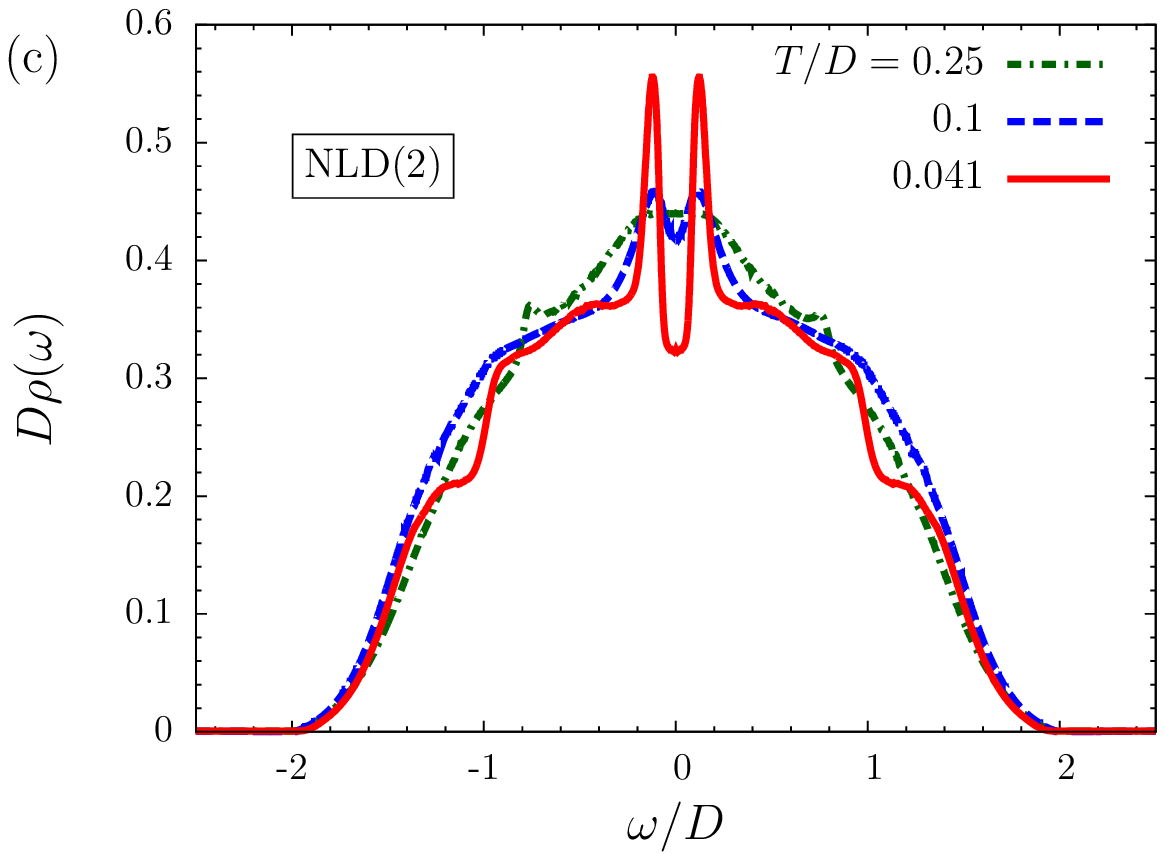}  
  \caption{(Color online) Comparison of the spectral function of the two-dimensional ($d=2$)
    half-filled Hubbard model on a simple cubic lattice for various
    temperatures $T$ and for different levels of the nonlocal approximation:
    (a) DMFT, (b) \DMFTTS(1), (c) \DMFTTS(2).
    The parameters in all calculations are $U/D=1$ and
    $\epsilon=-\frac{U}{2}=-\frac D2$.   
    All curves are calculated with the two-impurity enhanced noncrossing  
    approximation as impurity solver.
    The inset in panel (a) shows the noninteracting  density of states.
  }  \label{fig:res1}
\end{figure}

We study the Hubbard model on a two-dimensional simple-cubic lattice
with nearest-neighbor hopping only.  The half-bandwidth 
$D=2dt=4t$ is used as unit of energy and we set $k_\mathrm{B}=c=\hbar=1$.
The noninteracting spectral function of this system 
is shown in the inset of Fig.~\ref{fig:res1}(a).   

In order to investigate the influence of the nonlocal correlations we 
consider various stages of the scheme differing in the maximum
distance $\|\vek a\|$  to be incorporated into the 
approximation. This distance is measured with a Manhattan 
metric indicated by $\|.\|$  reflecting the minimum number of elementary 
hoppings between the two sites.
For example, the approximation denoted with \DMFTTS(1) includes  
only two-impurity models of  nearest-neighbor lattice sites, that is,
the set of distance vectors is given by
$\vek a_1\in \{(\pm1,0)^T,(0,\pm1)^T\}$. Accordingly, in \DMFTTS(2)
this set is augmented with the four next-nearest neighbor distances, i.e.\
the set of vectors is $\vek a_2\in \{\vek a_1, (\pm1,\pm 1)^T,(\pm 2,0)^T,(0,\pm 2)^T\}$
and so on.

Figure \ref{fig:res1} displays the local single-particle spectral function
\begin{align}
  \rho(\omega)=-\frac 1\pi\mathrm{Im}\frac 1N\sum_{\vek k}G_{\vek k}(\omega+i0^+)  
\end{align}
for three different temperatures and various stages of the approximation.
Panel (a) shows the usual DMFT solution, while panels (b) and (c) show 
the \DMFTTS(1) and  \DMFTTS(2) spectral functions, respectively.   
The sequence of decreasing temperatures reveals the development
of the well known many-body resonance at the Fermi level $\mu=\om=0$
in DMFT [panel (a)] indicating the formation of  low-energy 
quasiparticles.\cite{grewe:quasipart05} 

The inclusion of nonlocal correlations by utilizing \DMFTTS(1) 
leads to the formation of a pseudogap inside this resonance [see panels (b)].
A further inclusion of next-nearest neighbor correlations [panel (c)]
the gap widens and the sidepeaks become  more pronounced.
We cannot definitely decide whether or not a complete gap forms
at zero temperature ($T\rightarrow 0$)
where the  spectral function vanishes at the Fermi energy,
as it was found 
in a recent two-site DCA calculation.\cite{pruschke:nonlocalFluc2dHM05}
Too low temperatures can not be investigated with our impurity solver
due to the violation of Fermi liquid properties.\cite{schmittSus09,jabbenENCATI12}

Additionally, in the high-energy part of the spectrum the inclusion of
additional neighbors seems to bring out more of the van-Hove singularities
at $\om=\pm D$ of the original unperturbed ($U=0$) spectrum 
[see inset of Fig.~\ref{fig:res1}(a)].

\begin{figure}[ht]
  \includegraphics[width=0.9\linewidth]{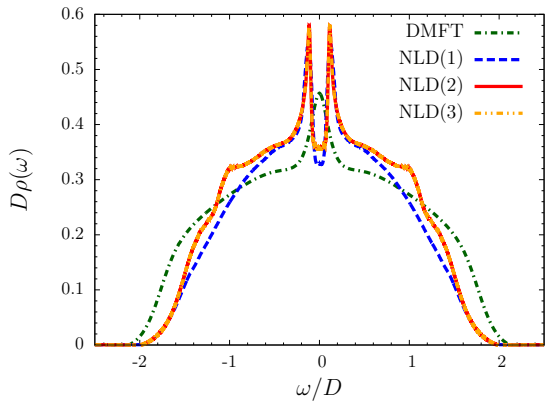}
  \caption{(Color online) Comparison of the spectral function for the different 
    levels of the \DMFTTS(i) approximation. The two-site correlations included 
    range from between 
    nearest neighbor sites in \DMFTTS(1) up to all pairs of 
    sites reachable by three elementary hoppings \DMFTTS(3).
    Other parameters are as in Fig.~\ref{fig:res1}.
  }  \label{fig:res2}
\end{figure}
Inclusion of additional  sites with $\|\vek a\|>2$ into our \DMFTTS-scheme does not
produce considerable changes, at least for this choice of parameters. 
This can be observed in Fig.~\ref{fig:res2}, where
spectral functions for a fixed temperature $T=0.05D$ 
are shown for various maximal distances up to $\|\vek a\| =3$,
i.e.\ including neighbors reachable by up to  three transfer
processes $\sim t$.  The changes become successively 
smaller and the curves from \DMFTTS(2) and \DMFTTS(3) are already almost 
indistinguishable.

Our calculations clearly show the appearance of high sidepeaks at the borders of
the pseudogap. The existence of sidepeaks in the low-energy spectrum is
in accord with numerical DCA results,\cite{huscroft:pseudogapHM01,*moukouriMIT01,*jarrell:QMCMEMNonLocalDMFT01}
where, however, large cluster sizes where necessary to resolve gap and
sidepeaks.

We would like to point out, that the sidepeaks in the low-lying
quasiparticle regime bear a strong resemblance to what is observed 
in  a related work\cite{jabbenENCATI12} for the TIAM.
In our opinion we now find a coherent version of the
splitting-scenario described there: The direct and the  induced 
effective hopping generate bonding and antibonding molecular-like orbitals
amongst different sites.  
The strength of this   strongly depends on the crystal structure since
the  relevant effective matrix elements vary and oscillate 
with distance.\cite{jabbenENCATI12} They  are also very sensitive  to the position of the Fermi 
level, which sets a scale for such oscillations.
The phase-information and coherence of electron propagations which is necessary for
such a splitting is fully incorporated into
our approach according to Eq.~\eqref{eq:2impT}.

These aspects also manifest themselves in the  sharpening of the
high-energy features around $\omega\approx D$ 
when going from \DMFTTS(1) to
\DMFTTS(2,3).

In addition to the bonding-antibonding splitting, an effective antiferromagnetic 
exchange $J_t=\frac{4t^2}{U}$ between neighboring sites is
generated which 
leads to the suppression of Kondo-like correlations.
Both effects described above
 induce a pseudogap and can lead to a splitting of the 
coherent many-body resonance at the Fermi level.
These cases could in principle be distinguished as 
the  splitting due to molecular binding should be linear in $|t|$, while
the magnetic singlet-triplet splitting  $J_t$ is proportional to $t^2$.
However, our choice of parameter values implies similar magnitudes 
for the hopping $t$ and the effective exchange $J_t$, 
and both effects should be of comparable size here. 

\begin{figure}[ht]
  \includegraphics[width=0.9\linewidth]{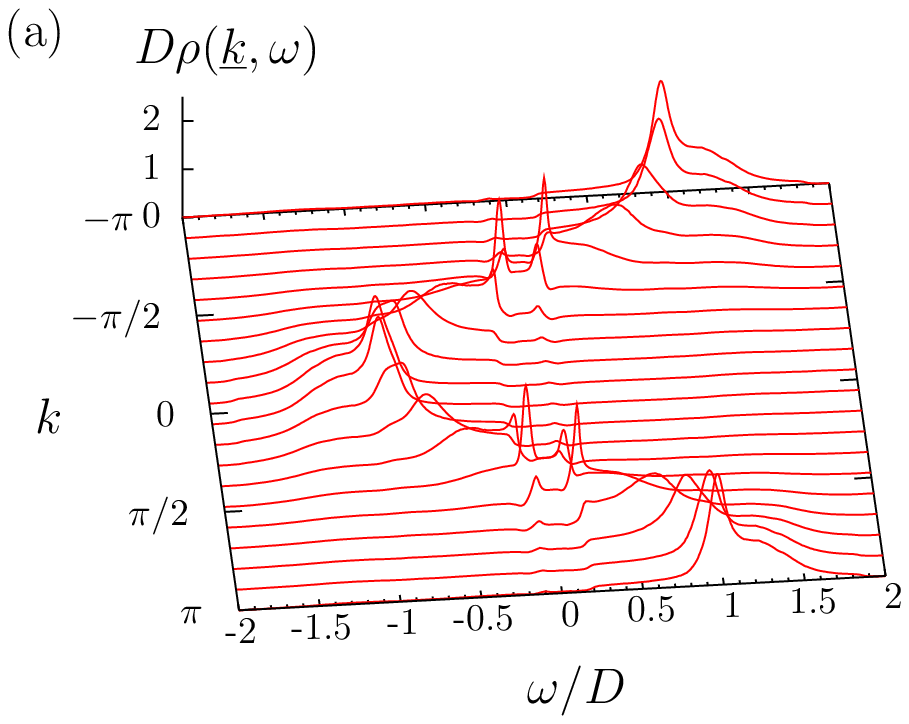}
  \includegraphics[width=0.9\linewidth]{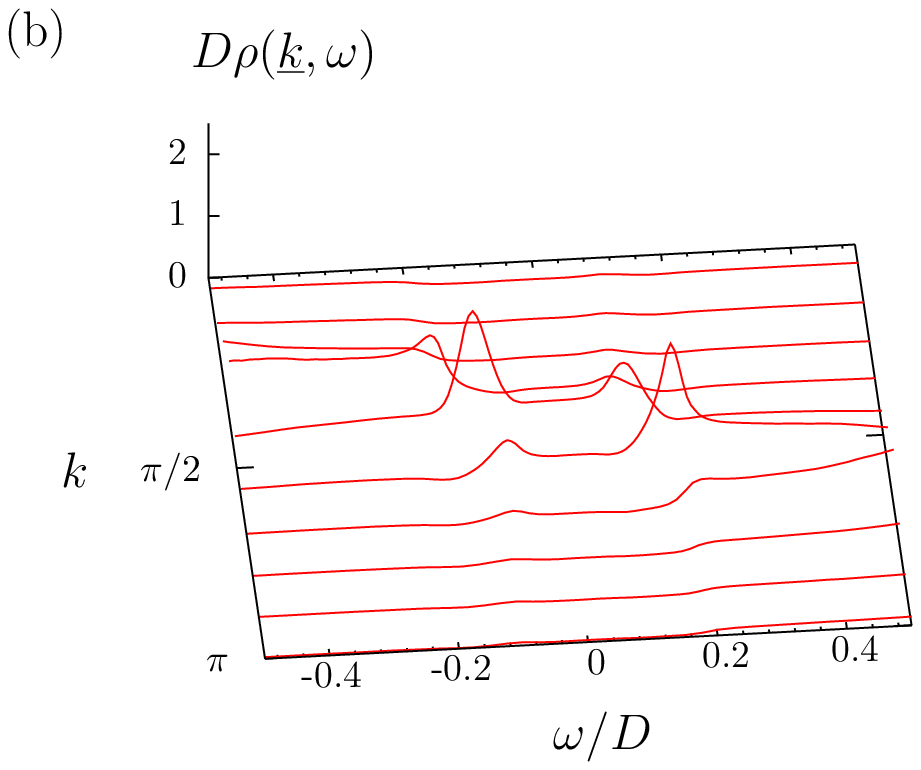}
  \caption{(Color online) $\vek{k}$-resolved spectral function obtained with the 
    \DMFTTS(2) for wave vectors along the 
    $(1,1)$-direction,  and for $\epsilon/D=-\frac 12$, $U/D=1$ and $T/D=0.05$. Panel (a) 
    shows the full energy interval while panel (b) displays a 
    magnification of the energy interval around  the Fermi-energy.
  }  \label{fig:res3}
\end{figure}
An interesting question is, whether the  excitations in  the vicinity of the
Fermi surface form a dispersive 
band or are narrowly concentrated in $\vek{k}$-space.
We show in Fig.~\ref{fig:res3}   the $\vk$-resolved spectral function
\begin{align}
  \rho(\vek k,\omega)=-\frac 1\pi\mathrm{Im}G_{\vek k}(\omega+i0^+)  
\end{align} 
for $\vek k$ along the diagonal
$(1,1)$-direction in the Brillouin zone. 
One still recognizes  the original cosine-form of
a tight-binding band via the positions 
of the peaks in the projection onto the $k-\om$-plane. 
However, these peaks develop a considerable width with growing $|\omega|$,  which indicates that
quasiparticle excitations are not well defined away from the Fermi
surface. Near the Fermi level $\om=0$ two distinct pairs of 
narrow peaks are visible at wave-vectors $\vk\approx \pm \frac{\pi}{2}(1,1)^T$.
Even though very slight remnants of these peaks are observable at neighboring wave vectors too,
their height is rapidly suppressed as $\vek k$ moves away from $ \pm \frac{\pi}{2}(1,1)^T$.
Since these maxima apparently furnish the spectral weight of the  flanks of the pseudogap
observable, e.g., in Fig~\ref{fig:res2}, 
we conclude those side-peaks to be rather localized in $\vek k $-space. Thus, 
they are
a result of increasing lattice coherence at decreasing temperature.  

Remarkably, also near the band edges at $\vek k=\vek 0$ and $\vek k=\pm \pi(1,1)^T$ 
with respective energies $\om\approx\pm D$ rather narrow peaks appear on top of the broad
resonances connected with the original band. Since they show up only
if next-nearest neighbors  are included, i.e. in \DMFTTS(2) but
not in \DMFTTS(1) (and DMFT), we attribute them to next-nearest neighbor
correlations, possibly of magnetic nature. A participation of an
indirect binding effect due to the repeated action of the transfer $t$,
i.e.\ a molecular orbital-like effect,
can also not be excluded with our choice of parameters.

Up to now a symmetric situation $2\eps+U=0$ was investigated, in
which a possible Mott-Hubbard gap as well as a possible pseudogap in
the low-energy quasiparticle regime both were to open around the
Fermi-level. Both effects can  be separated and identified individually
by moving away from 
half-filling, i.e.\ with  increasing doping.
Equivalently, we increase  the Coulomb-repulsion $U$ beyond the value $U=-2\eps$ 
and keep all other parameters fixed.

\begin{figure}[ht]
  \includegraphics[width=0.9\linewidth]{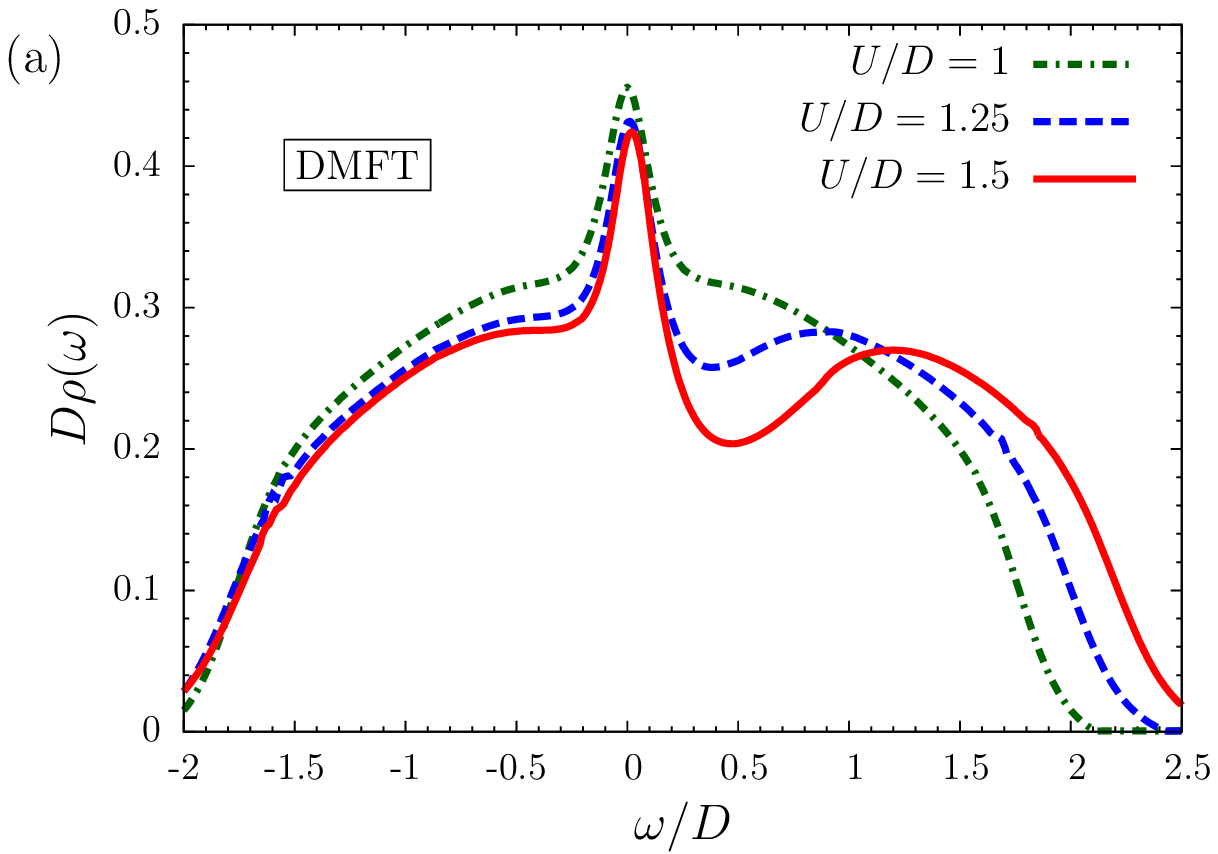}
  \includegraphics[width=0.9\linewidth]{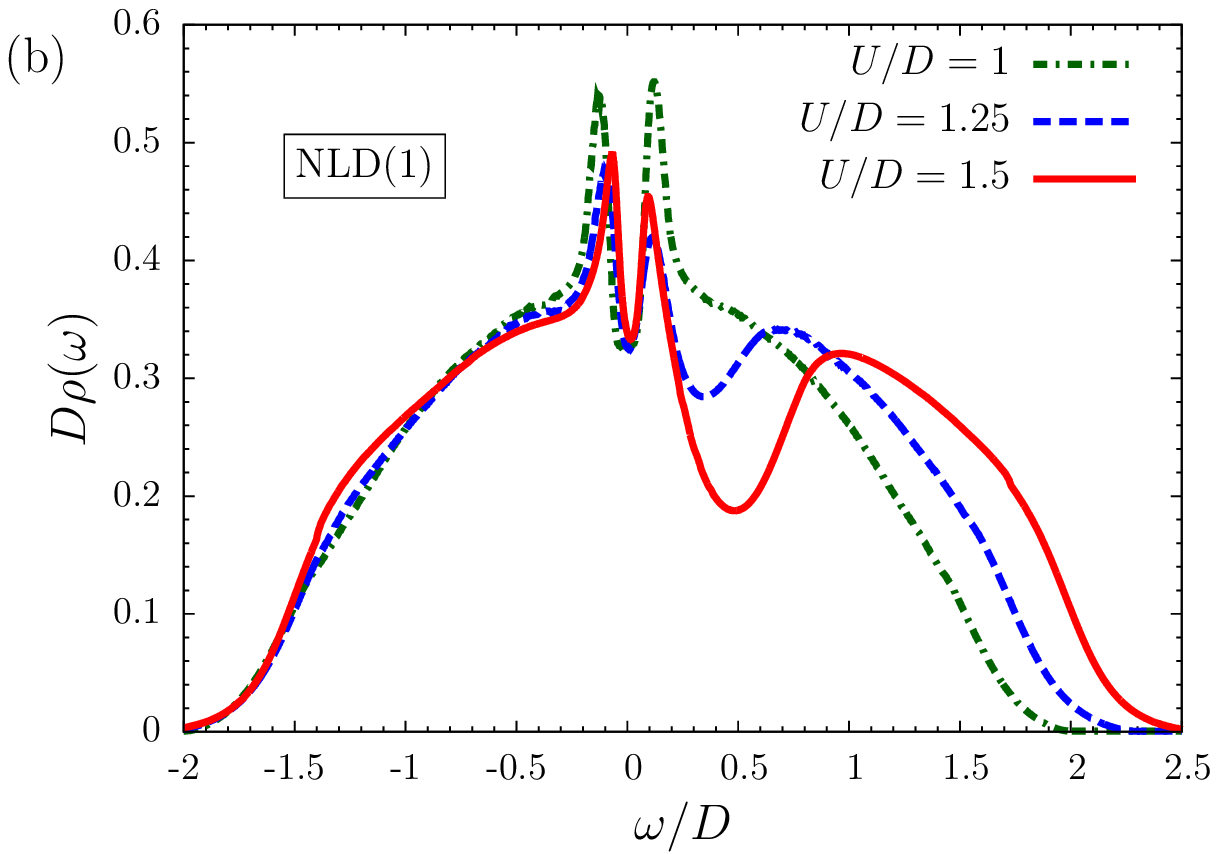}
  \includegraphics[width=0.9\linewidth]{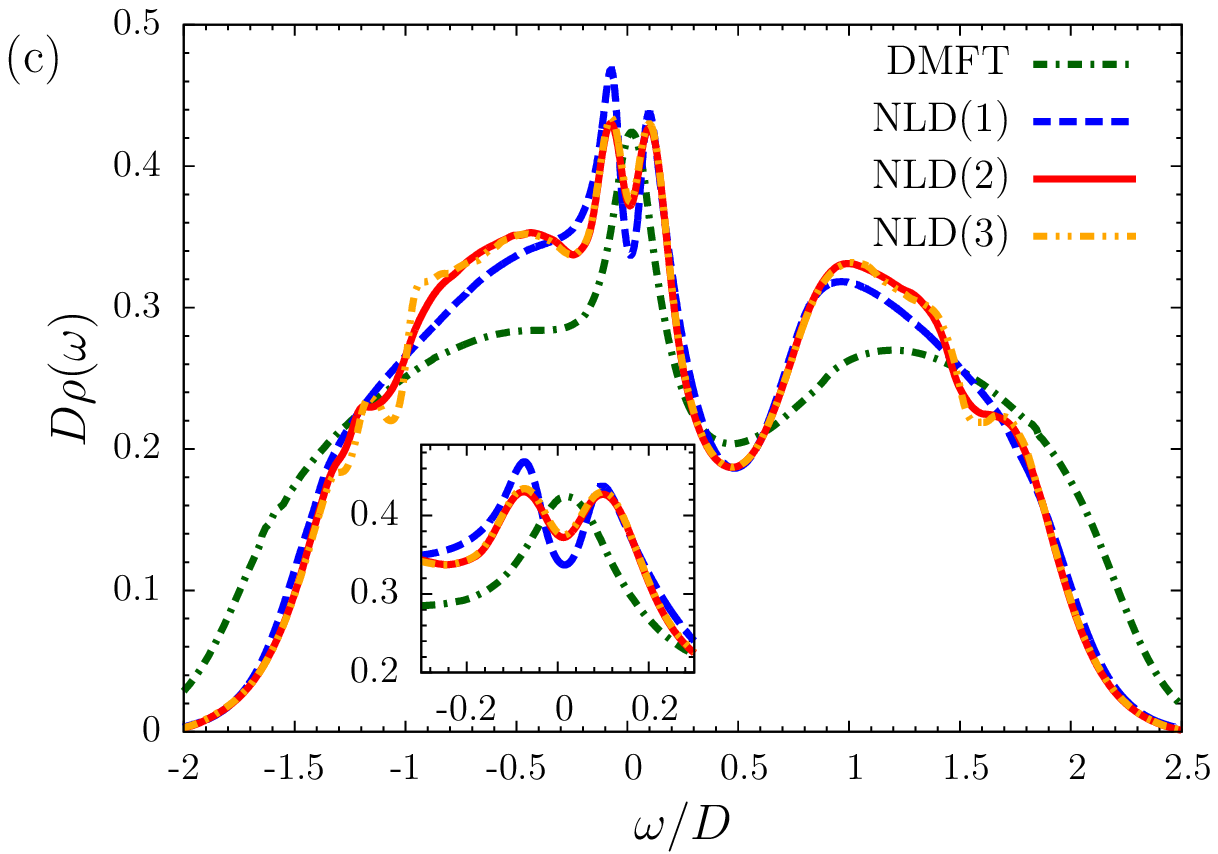}
  \caption{(Color online) Spectral functions for fixed $\epsilon=-\frac D2$
    and temperature $T/D=0.05$ for various $U$. (a) DMFT  (b) \DMFTTS(1). 
    (c) Spectral function for fixed $U/D=1.5$ and different stages of the 
    approximation as indicated. The inset shows a close up of the low-energy region
    around the Fermi level. The fillings corresponding to the curves shown in (c)    
    are $n_{\mathrm{DMFT}}\approx 0.90$ and $n_{\mathrm{\DMFTTS(i)}}\approx 0.92$. 
  }  \label{fig:res4}
\end{figure}
Calculations for such particle-hole asymmetric situations are shown in Fig.~\ref{fig:res4}. 
Panel (a) displays the formation of a Mott-Hubbard gap at  positive energies with increasing $U$
as calculated with the DMFT-approximation.
As expected, the center of the incipient gap moves away from the Fermi level
by an amount proportional to $U$  while the many-body resonance remains pinned at the Fermi level. 

Figure \ref{fig:res4}(b) shows results of \DMFTTS(1)-calculations
for the same  values of $U$. The Mott-Hubbard gap forms at positive energies
in a  similar fashion as in the DMFT-results of panel (a).  
Additionally, a pseudogap emerges in the many-body resonance at the Fermi level
as it was already found in earlier work.\cite{vidhyadhiraja2DHMNFL09,*khatamiQCPHM10,*yangQCSC11,chenVanHove11,macridinPseudogapHM06,linCDMFT10,*stanescuCDMFTMIT06,*maier:NCAIntersiteCorr00,huscroft:pseudogapHM01,*moukouriMIT01,*jarrell:QMCMEMNonLocalDMFT01,vilk:ShadowCuprate97,*Scalapino2006,*kyungPseudogap06,*tremblayPseudogap06}

The last part (c) of Fig.~\ref{fig:res4} shows the effect
of including more neighbors in our \DMFTTS(i)-scheme
for fixed Coulomb interaction $U$.
One again recognizes the dominant role of nearest-neighbor correlations 
since  the spectra with $i\geq 1$  all differ considerably from the DMFT-result. 
The low-energy region seems to be converged  for $i=2$, 
and the inclusion of next-nearest neighbors  
apparently  smoothens  the spikes around the pseudogap. 

In contrast, the inclusion of
next-next-nearest  neighbors ($i=3$) still 
has considerable impact on the high-energy features
at negative energies while at positive energies only 
small differences  are induced. 
This apparently indicates a reduced hybridization and corresponding 
band narrowing for states in the vicinity of the negative band edge. 
The importance of lattice structure, 
coherence and correlations for the possible development of hybridization
gaps and band splittings as they were already discussed in the 
symmetric case is evident from these results.

The longer-ranged correlations  also seem to work in favor of restoring particle-hole symmetry 
in the low-energy region [see inset of Fig.~\ref{fig:res4}(c)],
a phenomenon which is also discussed in connection with the cuprate 
superconductors.\cite{vidhyadhiraja2DHMNFL09,*khatamiQCPHM10,*yangQCSC11,chenVanHove11,chakrabortyPHSymm10}

\begin{figure}[ht]
  \includegraphics[width=0.9\linewidth]{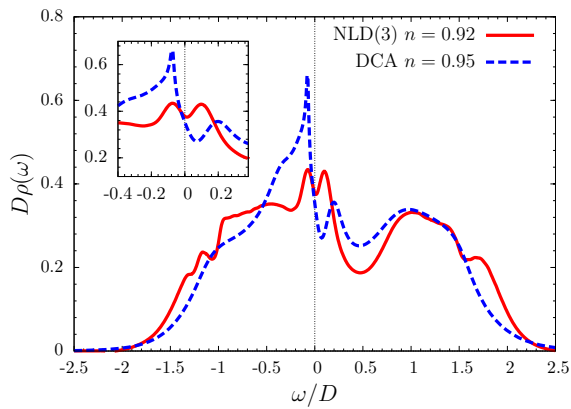}
  \caption{(Color online) Comparison of  \DMFTTS(3) and DCA spectral functions.
    The  \DMFTTS(3) curve is the one shown in Fig.~\ref{fig:res4}(c).
    The DCA result is taken from Fig.~5 of Ref.~\onlinecite{chenVanHove11}
    and 
    was obtained from self-consistent  quantum Monte Carlo solution of a 16-site cluster
    for $U/D=1.5$ and $T/D=0.017$.
    The filling for the DCA calculation of 
    $n_{\mathrm{DCA}}\approx 0.95$  is slightly larger 
    than the \DMFTTS(3) filling, $n_{\mathrm{\DMFTTS(3)}}\approx 0.92$.
    The inset shows an enlargement of the region around the Fermi level. 
  }  \label{fig:res5}
\end{figure}
Figure \ref{fig:res5}  compares  a \DMFTTS(3) spectral function
for finite doping and  $U=1.5D$  with
a spectrum obtained from a DCA calculation as it was
 published in Fig.~5 of  Ref.~\onlinecite{chenVanHove11}. 
The qualitative features of both approaches
nicely agree: the humps and minima visible in the DCA spectra translate to
corresponding but more pronounced features in the \DMFTTS(3). 
Given the different nature of the two approximations, this provides additional 
evidence for the physical nature of the observed structures.

The magnitude of the pseudogap is of the same order in both approaches,
although the DCA result is slightly larger and the low-energy
spectrum much more asymmetric.
One reason for this can be found in the enhanced two-impurity noncrossing approximation
as the two-impurity solver, which is known 
to produce a too small  many-body low-energy scale for the SIAM. This
is especially relevant for the low-energy region and 
also translates to self-consistent approximations 
(see, e.g.\ Ref.~\onlinecite{greteKink11}).

The different temperatures of both calculations should not be relevant for
the qualitative features as $T$ is already rather low and
further decreasing  it will only slightly deepen the
pseudogap.     

However, a qualitative difference between the two calculations  is found 
in the different fillings 
which in general strongly influences the low-energy spectral function.
%
The analysis in Ref.~\onlinecite{chenVanHove11} revealed that decreasing the filling 
from $n_{\mathrm{DCA}}\approx 0.95$ to $n_{\mathrm{DCA}}\approx 0.88$ (for otherwise identical parameters)
leads to a spectral function which is nearly particle-hole symmetric at low
energies and does not exhibit a pseudogap. In this light, the  \DMFTTS(3)
spectral function for $n_{\mathrm{\DMFTTS(3)}}\approx0.92$ 
with a nearly symmetric low-energy  spectrum and a small  pseudogap
represent a plausible intermediate solution.

\section{Conclusion}
\label{sec:conclusion}
We have proposed a novel general and nonperturbative 
scheme for the treatment of correlated electron systems 
on crystal lattices. Our approach
represents an extension of the well-known DMFT to include
nonlocal two-site correlations with arbitrary spatial extent. 

The self-consistent formulation establishes a mapping between nonlocal
two-site correlations of the lattice model and the equivalent 
functions of various two-impurity Anderson models with varying distance.
The two-impurity models are 
solved in real-space, but the extracted correlation functions
are transformed into momentum space via a Fourier transform.
Thereby translational invariance is built into our scheme  
by construction.

Most important, the spatial range of correlations which are explicitly 
included in the treatment is  unrestricted in principle.
It corresponds to the maximum impurity-impurity distance in an 
effective TIAM which is solved in our approach, and is therefore 
only limited by the accuracy of the two-impurity solver. 

As a first application, we  applied  our scheme  to the 
Hubbard model on a two-dimensional simple-cubic lattice.
We found that  the leading nonlocal correlations produce a
pseudogap in the low-energy single-particle excitation spectrum
as it is to be expected.\cite{vidhyadhiraja2DHMNFL09,*khatamiQCPHM10,*yangQCSC11,chenVanHove11,macridinPseudogapHM06,linCDMFT10,*stanescuCDMFTMIT06,*maier:NCAIntersiteCorr00,huscroft:pseudogapHM01,*moukouriMIT01,*jarrell:QMCMEMNonLocalDMFT01,vilk:ShadowCuprate97,*Scalapino2006,*kyungPseudogap06} 
Moreover, pronounced side-peaks occur as  a signature of increasing coherence.
In a situation without particle-hole symmetry we could clearly discriminate 
between the  Mott-Hubbard gap induced by mostly local atomic  correlations 
and the pseudogap, which is found in the many-body resonance at the Fermi level.  
We have also demonstrated that the inclusion of correlations over larger distances
brings out more details of the excitation spectra in the high-energy region.

Additionally, our method compares very well to results from the DCA. 
The spectral functions of both approaches exhibit similar qualitative 
features for finite doping, although they are more pronounced 
in our scheme. Also, the pseudogap obtained within both approaches 
is of the same  magnitude.

Thus, the scheme presented  opens an excellent
perspective for more detailed investigations of systems where nonlocal correlations 
play an important role.
These include phase transitions and critical phenomena in general, and 
quantum  critical points in particular,  as they are found, for example,
in heavy fermion compounds
or cuprate superconductors.

Also, the calculation of nontrivial critical exponents and their
scaling behavior\cite{rohringer3dHM11} seems to be in reach using  this new approach.  
In some cases it will nevertheless be necessary to include higher-order irreducible correlations 
beyond those between only  two sites.  This will be of particular importance,  when  complicated ground 
states involving correlations on plaquettes of sites or general resonating 
valence bond-states are under consideration.\cite{PhysRevLett.105.267204,*nature08942,*1173.full}

\begin{acknowledgments}
  We thank Eberhard Jakobi for fruitful discussions, Kuang-Shing Chen for providing us with
  the DCA data shown in Fig.~\ref{fig:res5},
  and the NIC, Forschungszentrum J\"ulich,  for their supercomputer support under 
  Project No. HDO00. 
  SS acknowledges financial support from the Deutsche Forschungsgemeinschaft 
  under Grant No. AN 275/6-2. 
\end{acknowledgments}


 


\end{document}